\documentstyle[12pt]{article}
\newcommand{\p}[1]{(\ref{#1})}
\newcommand{\be}{\begin{equation}}
\newcommand{\bea}{\begin{eqnarray}}
\newcommand{\ee}{\end{equation}}
\newcommand{\eea}{\end{eqnarray}}
\newcommand{\cp}{\mbox{$\cal P$}}
\newcommand{\e}{\eta}
\newcommand{\la}[1]{\langle S_{#1}| }
\newcommand{\ra}[1]{|S_{#1}\rangle }

\begin{document}
\topmargin -1cm
\oddsidemargin=0.25cm\evensidemargin=0.25cm
\setcounter{page}0
\renewcommand{\thefootnote}{\fnsymbol{footnote}}
\thispagestyle{empty}
{\hfill  Preprint JINR E2-97-63}\vspace{1.5cm} \\
\begin{center}
{\large\bf  Dimensional reduction and BRST approach to the description 
of a Regge trajectory
}\vspace{0.5cm} \\
A. Pashnev\footnote{E-mail: pashnev@thsun1.jinr.dubna.su}\\
and M. Tsulaia\footnote{E-mail: tsulaia@thsun1.jinr.dubna.su}
\vspace{0.5cm} \\
{\it JINR--Bogoliubov Laboratory of Theoretical Physics,         \\
141980 Dubna, Moscow Region, Russia} \vspace{1.5cm} \\
{\bf Abstract}
\end{center}
\vspace{1cm}

The local free field theory for Regge trajectory 
is described in the framework
of the BRST - quantization method. The corresponding BRST - charge
is constructed with the help of the method of dimensional reduction.

\vspace{0.5cm}
\begin{center}
{\it Submitted to Modern Physics Letters A}
\end{center}

\newpage\renewcommand{\thefootnote}{\arabic{footnote}}
\setcounter{footnote}0\setcounter{equation}0
\section{Introduction}

The main problem in the description of the higher spin particles
is removing of unphysical degrees of freedom from the theory.
As was shown in \cite{FP}, the corresponding lagrangian must have
some invariance, which generalizes the gauge invariance of the
electromagnetic field not only in the free case, but for interacting
fields, as well.

On the free level such
 lagrangians was constructed both for massive
 and for massless particles 
of any spin as well as for massless supermultiplets \cite{Fr}-\cite{AD}.
In some sense the massless case is simpler and, hence, more investigated
than the massive one. 
The hope is, that with the help of some Higgs-type effect some
of the interacting particles  acquire nonzero values of the mass.

In general the lagrangians for higher spins include additional fields. 
Some of them are auxiliary, others can be gauged away.
The main role of these fields is to ensure, that only physical degrees
of freedom are propagating ones. It means, that the basic field
describes irreducible representation of the Poincare group - it
is traceless, transverse and satisfies mass shell equation.

Naturally higher spin particles arise after quantization of classical
extended objects such as string, relativistic oscillator \cite{K}-
\cite{BD},
discrete string \cite{GP} etc. Physically they correspond 
to exited levels
of the system and belong to Regge trajectories, each including
infinite sequence of states with a spin linearly depending on the
square of the mass. There are infinite number of Regge trajectories
in the string and relativistic oscillator models and only one such
trajectory in the discrete string model due to existing of additional
second-class constraints. Such subdivision of all higher spin particles
on Regge trajectories leads to the consideration 
of these Regge trajectories
as independent objects for which it would be interesting to construct
the lagrangian description.

One of the most economic and straightforward method of construction of
lagrangians for systems with constraints 
is the method using BRST-charge of the
corresponding firstly quantized theory. With the help of BRST-charge
the lagrangians for the free field theory for particular values of spins
\cite{M} and for infinite tower of massless
higher spin particles were constructed \cite{OS}. 
The analogous consideration for the massive fields is hampered by the
presence of the second class constraints. 
The methods of converting them into the first class constraints 
with the help of additional variables 
were discussed in \cite{FS}-\cite{EM}.
Nevertheless, the strightforward application of this
approach to the massive higher spins leads to the non-local lagrangian
\cite{PT}.

In the second part of the paper we describe the auxiliary Fock space and
system of constraints in it which follow from the consideration of
some classical models of extended objects.

In the third part of the paper 
we describe the massless case of \cite{OS} in $D+1$ dimensions
which is the starting point for the consequent consideration. 
After that  we fulfil the dimensional reduction
from ${D+1}$ dimensional to $D$ dimensional space time.
Formally it looks like some unitary transformation.
As a result, the correct local free field lagrangian
for Regge trajectory together with its daughter trajectories is
constructed.

In the fourth part of the paper we write out as examples 
the lagrangians for lower spins $S=0,1,2$.

\section{Constraints in auxiliary space}

To describe all higher spins simultaneously it is convenient to
introduce auxiliary Fock space generated by creation and annihilation
operators $a_\mu^+,a_\mu$
with vector Lor
entz index $\mu =0,1,2,...D-1$, satisfying the following
commutation relations
\be
[ a_\mu,a_\nu^+ ] =-g_{\mu \nu},\;g_{\mu \nu}=diag(1,-1,-1,...,-1).
\ee
In addition the operators $a_\mu^+,a_\mu$  can have some internal
indices leading to more complicated spectrum of physical states.
For simplicity we consider in this paper $a_\mu^+,a_\mu$
without additional indices.

The general state of the Fock space
\be
|\Phi\rangle =\sum \Phi^{(n)}_{\mu_1\mu_2\cdots\mu_n}(x)a_{\mu_1}^{+}
a_{\mu_2}^+\cdots a_{\mu_n}^+ |0\rangle
\ee
depends on space-time coordinates $x_\mu$ and its components
$\Phi^{(n)}_{\mu_1\mu_2\cdots\mu_n}(x)$ are tensor fields of rank $n$
in the space-time of arbitrary dimension $D$.
The norm of states in this Fock space is not positively definite due to
the minus sign in the commutation relation (2.1) for time components of
creation and annihilation operators. It means that physical states must
satisfy some constraints to have positive norm. These constraints
arise naturally in the considerations of classical composite systems
\cite{BD}-\cite{GP}.
The corresponding quantum operators
\bea \label{GEN}
L_0 &=&-{p_{\mu}}^2-\alpha'a_\mu^+a_\mu,\;\;\;
L_1 = p_{\mu}a_{\mu},\;\;\; L_{-1}=p_{\mu}a_{\mu}^+ =L_1^+,\\ \label{l2}
L_2 &=& \frac{1}{2}a_{\mu}a_{\mu},\;\;\; L_{-2}=\frac{1}{2}
a_{\mu}^+a_{\mu}^+=L_2^+
\eea
form the algebra
\bea
[L_0\;,\;L_{\pm 1}]=\mp\alpha'L_{\pm 1} & [L_0\;,\;L_{\pm 2}]= &
\mp 2\alpha'L_{\pm 2}\\
\eea
\bea
[L_{1}\;,\;L_{-2}]=-L_{-1} & [L_{-1}\;,\;L_2]= & L_{1}\\
\eea
\bea
[L_1\;,\;L_{-1}]=-{p_{\mu}}^2 & [L_2\;.\;L_{-2}] =&-a_\mu^+a_\mu+
\frac{D}{2}
\equiv G_0
\eea
The operators $L_0, L_1, L_2$ correspond to the mass shell,
transversality and tracelessness conditions on the wavefunctions.
The operators $L_1 , L_{-1}$ and  $L_2 , L_{-2}$ are of second class
and in general this system of constraints describes single Regge
trajectory \cite{BD}-\cite{P}.

There exist some different possibilities in consideration of
general system of constraints \p{GEN}-\p{l2}.
 The truncated system $L_0 , L_{\pm 1}$
describes Regge trajectory together with its
daughter trajectories. The operators $L_{\pm 1}$ in this system
are of second class as before. We describe the BRST - quantization
of this system in the third part of the article.

The limit $\alpha'=0$ of the system \p{GEN}
corresponds to the massless infinite tower
of spins with a single state  at each value of the spin.
In this case only the operators $L_{\pm 2}$ are of second class.
The BRST - quantization of this system as well as general system
\p{GEN}-\p{l2} will be given elsewhere.

The simplest system of first-class constraints
 \begin{equation}\label{OS}
\tilde{L}_0=-{p_{\mu}}^2\;,\;L_{\pm 1}
\end{equation}
corresponds to the massless tower of spins infinitely degenerated
at each value of spin. The BRST - construction for this system of
constraints was described in \cite{OS}-\cite{M}.

\setcounter{equation}0\section{Massless case and dimensional reduction}

In this section we consider the system with constraints
\begin{equation}
L_0 =-{p_{\mu}}^2-\alpha'a_\mu^+a_\mu+\alpha_0,\;\;\;
L_1 = p_{\mu}a_{\mu},\;\;\; L_{-1}=p_{\mu}a_{\mu}^+,
\end{equation}
where parameter $\alpha_0$
 plays the role of intercept
for Regge trajectory.
 In some sense this system is
intermediate  between the systems in \cite{OS} and \cite{P} because
it describes the Regge trajectory together with all daughter trajectories
(due to the absence of the constraints $L_{\pm 2}$).
The commutation relation $[L_1\;,\;L_{-1}]=-{p_{\mu}}^2 $ means that
$L_{\pm 1}$ are the second class   constraints . To construct the 
BRST - charge we can try to convert them into the first class constraints
following the prescription of \cite{FS} - \cite{EM}. We 
introduce  new operators $b$ and $b^+$ with the commutation relations
$[b,b^+]=1$ and modify  the constraints to the following expressions:
\begin{eqnarray}            \label{AAA}
\tilde L_0& = &-{p_{\mu}}^2
-\alpha'a_\mu^+a_\mu+\alpha_0 + \alpha'b^+b,\\
\tilde L_{-1}& = &L_{- 1} + \sqrt{{p_{\mu}}^2}b^+,\\
\tilde L_{1}& = &L_{1} + \sqrt{{p_{\mu}}^2}b,
\end{eqnarray}

\begin{equation}\label{A1}
[\tilde L_1 , \tilde L_{-1}]=0, \quad [\tilde L_0,\tilde L_1] =
- \alpha'\tilde L_1,   \quad [\tilde L_0,\tilde L_{- 1}] =
\alpha'\tilde L_{-1}.
\end{equation}

All of the modified constraints are of first class and BRST - charge
construction is straightforward \cite{PT}. Nevertheless, the corrresponding
lagrangian is not satisfactory one. Indeed, it is non-local due to
presence of $\sqrt{{p_{\mu}}^2}$ in the definitions of constraints
$\tilde L_{\pm 1}$. The natural way out of this difficulty is to
replace $\sqrt{{p_{\mu}}^2}$ by 
$\sqrt{-\alpha'a_\mu^+a_\mu+\alpha_0 + \alpha'b^+b},$
having new system of constraints 
with rather nontrivial commutation relations due to such complicated
dependence from creation and annihilation operators.

The simplest way to construct BRST - charge with 
corresponding constraints
is as follows. Firstly we consider massless case in $D+1$ dimensions
with constraints
\begin{eqnarray}
L_0&=&-p_{\mu}^2+p_D^2, \;\;\;\;\mu=0,1,...,D-1,\\
L_1&=&p_\mu a_\mu-p_Da_D,\\
L_{-1}&=&p_\mu a_\mu^+ -p_Da_D^+.
\end{eqnarray}
Following to the standard procedure we
introduce additional set of anticommuting variables
$\e_0,\e_1,\e_1^+$ having ghost number one and corresponding momenta
$\cp_0,\cp_1^+,\cp_1$ with commutation relations:
\begin{equation}
\{\e_0,\cp_0\}=\{\e_1,\cp_1^+\}=\{\e_1^+,\cp_1\}=1.
\end{equation}
The nilpotent BRST - charge has the following form:
\begin{equation}
Q =\e_0  L_0 + \e^+_1 L_1 + \e_1  L_{1}^+  +
\e_1 \e^+_1 \cp_0
\end{equation}
Consider the total Fock space generated by creation operators
$a_\mu^+,a_D^+,\e_1^+,\cp_1^+$. In addition each vector of the Fock
space depends linearly on the real grassmann variable $\e_0$
($\cp_0$ considered as corresponding derivative
$\cp_0=\partial / \partial\e_0$)
\begin{equation}
|\chi\rangle =  |\chi_1\rangle +\e_0 |\chi_2\rangle.
\end{equation}
Ghost numbers of $|\chi_1\rangle$ and $|\chi_2\rangle$ are different if
the state $|\chi\rangle$ has some definite one.

The BRST - invariant lagrangian in such Fock space can be written as
\begin{equation} \label{L11}
L=-\int d \e_0 \langle\chi|Q|\chi\rangle.
\end{equation}
To be physical, lagrangian $L$ must have zero ghost number. It means
that vectors $|\chi\rangle$ and $|\chi_1\rangle$  have zero ghost
numbers as well. In this case the ghost number of $|\chi_2\rangle$
is minus one. The most general expressions for such vectors are
\begin{eqnarray}
|\chi_1\rangle&=&\ra{1}+\e_1^+\cp_1^+\ra{2},\\
|\chi_2\rangle&=&\cp_1^+\ra{3},
\end{eqnarray}
with vectors $\ra{i}$ having ghost number zero and depending only
on bosonic creation operators $a_\mu^+,a_D^+$
\begin{equation}\label{RA}
\ra{i}=\sum \phi^{n}_{\mu_1,\mu_2,...\mu_n}(x)
a_{\mu_1}^+ a_{\mu_2}^+ ...a_{\mu_n}^+(a_D^+)^{n}|0\rangle.
\end{equation}

Integration over the $\e_0$ gives the following expression for the 
lagrangian in terms of $\ra{i}$
\begin{eqnarray}\label{LT}
L&=&\la{1}{p_{\mu}}^2 \ra{1} - \la{2}{p_{\mu}}^2\ra{2} - \la{3}\ra{3}+ \\
&&\la{1} L^+_{1}\ra{3} + \la{3} L_1 \ra{1} -
\la{2} L_1 \ra{3} - \la{3} L^+_{1} \ra{2}.\nonumber
\end{eqnarray}

The nilpotency of the BRST - charge leads to the invariance of the
lagrangian \p{L11} under the following transformations
\begin{equation}\label{TR}
\delta |\chi \rangle = Q|\Lambda \rangle.
\end{equation}
The parameter of transformation must have ghost number $-1$ and
can be written as  $|\Lambda \rangle = \cp^+_1|\lambda\rangle ,$
where $|\lambda\rangle$ belong to the Fock space generated by
$a_\mu^+ , a_D^+$ and depends from the space - time coordinates.

Using together this invariance and equations of motion
for the fields $\ra{i}$
one can show , that the lagrangian \p{LT} describes the 
infinite number of higher spin massless particles with infinite
multiplicity at each value of spin \cite{OS}. 

To obtain the description of the Regge trajectory in $D$ dimensions
one can make use of the dimensional reduction procedure to the $D+1$
dimensional massless lagrangian.
The masslessness condition in $D+1$ dimensions is
\begin{equation}
p_\mu^2-p_D^2=0.
\end{equation}
After the fixing $p_D=m$ with some arbitrary parameter $m$ 
this equation describes massive particle. In principle $m$ can
depend from the spin of particle, leading to the Regge trajectory.

To describe the linear Regge trajectory we fix the following $x_D$
dependence of the Fock space vector $| {\chi} \rangle $
in \p{L11}:
\begin{equation}               \label{sub}
|{\chi}\rangle =
 exp (ix_D \sqrt {{\alpha}'(-a^+_{\mu}a_{\mu} + a^+_D a_D  +
 \e^+_1 \cp_1 + \cp^+_1 \e_1) + \alpha_0}|\chi^{\prime}\rangle \equiv
U|\chi^{\prime}\rangle
\end{equation}
The result of the substitution of \p{sub} in the expression \p{L11}
is
\begin{equation} \label{L12}
L=-\int d \e_0 \langle {\chi}^{\prime} |\tilde{Q}| {\chi}^{\prime} 
\rangle,
\end{equation}
Where
\begin{eqnarray}\label{q}
\tilde Q&=&\e_0  (\tilde L_0 + {\alpha}'(\e^+_1 \cp_1 +
\cp^+_1 \e_1)) + \e_1^+(\tilde L_1 + \cp^+_1 X \e_1 a_D) +
\nonumber \\
&&+(\tilde L^+_1 + a^+_D  \e^+_1 X \cp_1 )\e_1 +
\e_1 \e^+_1 \cp_0
\end{eqnarray}
with the following notations used
\begin{equation}
\tilde L_0 = - {p_{\mu}}^2 + {\alpha}'(-a^+_{\mu}a_{\mu} + a^+_D a_D) +
\alpha_0
\end{equation}
\begin{equation}              \label{l1}
\tilde L_1 = p_{\mu}a_{\mu} - 
\sqrt{{\alpha}'(-a^+_{\mu}a_{\mu} + a^+_D a_D)+
{\alpha}' + \alpha_0}\;\; a_D
\end {equation}
\begin{equation}                       \label{l-1}
\tilde L^+_1 = p_{\mu}a^+_{\mu} -a^+_D
\sqrt{{\alpha}'(-a^+_{\mu}a_{\mu} + a^+_D a_D)+
{\alpha}' + \alpha_0}
\end {equation}
\begin{equation}
X=-\sqrt{{\alpha}'}
\left (\sqrt{-a^+_{\mu}a_{\mu} + a^+_D a_D + 
{\alpha_0\over{\alpha}'} + 2}-
\sqrt{-a^+_{\mu}a_{\mu} + a^+_D a_D + 
{\alpha_0 \over {\alpha}'} + 1} \right )
\end{equation}
The commutator of constraints \p{l1} and \p{l-1} has rather complicated
form:
\begin{equation}
[\tilde L_1,\tilde L_1^+] = \tilde L_0 + {\alpha}' + a_D^+X \tilde L_1  +
\tilde L_1^+ X a_D + a_D^+X^2 a_D.
\end{equation}
The new BRST - charge $\tilde{Q}$ \p{q} is nilpotent due to 
unitarity of the transformation $\tilde{Q}=U^{-1} Q U$. Our choice of 
square root dependence in the exponent in \p{sub} leads, as we will see
later, to the linear Regge trajectory. Replacement of this square root
by any other function consistently gives 
Regge trajectory with more complicated 
dependence between spin and mass.
The nilpotency of the BRST - charge evidently does not depend from the
choice of this function.

In terms of $\ra{i}$ the lagrangian \p{L12} has the folloqing form: 
\begin{eqnarray}\label{LT}
L&=&-\la{1}{\tilde L_0} \ra{1} + \la{2}(\tilde L_0 + 2 {\alpha}')\ra{2}
+\la{1} \tilde L^+_1 \ra{3} + \la{3} \tilde L_1 \ra{1} -
\nonumber \\
&&\la{2} (\tilde L_1 + X a_D)\ra{3} - \la{3} (\tilde L^+_1 +a^+_D X)\ra{2}
- \la{3} \ra{3}.
\end{eqnarray}
The correswponding equations of motion are:
\begin{eqnarray}
&&\tilde L_0 \ra{1}-\tilde L^+_1 \ra{3}=0,\\ \label{s2}
&&(\tilde L_0 +2\alpha')\ra{2}-(\tilde L_1+Xa_D) \ra{3}=0, \\
&&\ra{3}-\tilde L_1 \ra{1}+(\tilde L^+_1+a_D^+X) \ra{2}=0.
\end{eqnarray}
The transformation law $\delta {\mid} {\chi}' \rangle = 
\tilde Q{\mid}{\Lambda}{\rangle}$
has the following component form:
\begin{eqnarray}
\label{To1}
\delta{\ra{1}}&=&\tilde{L}_1^{+}{\mid}\lambda\rangle,\\ \label{To2}
\delta{\ra{2}}&=&({\tilde{L}_1} + X a_D){\mid}\lambda\rangle,\\ \label{To3}
\delta\ra{3}&=&(\tilde{L}_0 + {\alpha}'){\mid}\lambda\rangle.
\end{eqnarray}

One can show that using together \p{To1} and equations of motion
for the fields $\ra{i}$
one can eliminate the fields $\ra{2}$ and $\ra{3}$. Firstly we
solve the equation
\begin{equation}\label{S3}
\ra{3}+(\tilde L_0 + {\alpha}')|\lambda \rangle = 0
\end{equation}
using decompositions
\begin{equation}
|S_i\rangle =\sum (a_D^+)^n|S_{in}\rangle,\;\;
|\lambda\rangle =\sum (a_D^+)^n|\lambda_{n}\rangle.
\end{equation}
The equation \p{S3} does not fix parameter $|\lambda \rangle$ completely.
There will be residual invariance with parameter
$|{\lambda}^{\prime}\rangle$ under the condition
\begin{equation}\label{c1}
(\tilde L_0 + \alpha')|{\lambda}^{\prime}\rangle=0.
\end{equation}
After the elimination of the field $|S_2\rangle$
with the help of the equation $
\ra{2}+(\tilde L_1+Xa_D)|{\lambda}^{\prime}\rangle=0
$, which is consistent with the equations \p{s2} and \p{c1},
the new parameter
$|{\lambda}^{\prime\prime} \rangle$ will satisfy two conditions
$(\tilde L_0+\alpha')|{\lambda}^{\prime\prime}\rangle=
(\tilde L_1+Xa_D)|{\lambda}^{\prime\prime}\rangle=0$.
With the help of this parameter all fields $|S_{1n}\rangle$,
except $|S_{10}\rangle$ can be eliminated as well. 
It means that only $|S_{10}\rangle$ under conditions
\begin{equation}
L_0 \ra{1}=0, \quad L_1 \ra{1}=0
\end{equation}
is a physical field. The second  condition kills negative norm states
of the Fock space. The first one - the mass shell condition - fixes
the linear dependence between masses and spins of physical states.

In general   the Fock space vectors $|\chi\rangle$ and $|\chi'\rangle$
have complex vawefunctions. In the massless case one can impose 
the following reality conditions consistent with the equations of 
motion and transformation law \p{TR}: the coefficients in $\ra{1}$ and
$\ra{2}$ are real as opposite to the coefficients in $\ra{3}$ and 
$|\lambda\rangle$, which are purely imaginary. The transformation
\p{sub}, due to its complexity destroys such reality conditions for the 
massive case.
\setcounter{equation}0\section{Examples}
The total lagrangian \p{LT} describes all spins from zero to infinity.
The maximal spin at the level $m^2=\alpha_0+ n \alpha'$ is $n$. Due to
luck of the tracelessness constraint, there are also spins
$n-2,\; n-4\; ...$ on this level. The part of the lagrangian 
describing this level contains fields in $\ra{1}, \ra{2}$ and $\ra{3}$
with $n, n-2$ and $n-1$ total numbers of creation operators $a_\mu^+$
and $a_D^+$. In this chapter we describe three simplest cases 
$n=0, 1, 2$.

For $n=0$ the only field 
\begin{equation}
\ra{1}_0=A|0 \rangle
\end{equation}
gives contribution to the lagrangian: 
\begin{equation}
L=-A{{\partial}_{\mu}}^2A - \alpha_0 A^2
\end{equation}
Obviously it describes spinless particle with $m^2=\alpha_0$.

For $n=1$
corresponding fields are:
\begin{equation}
\ra{1}_1=(A_{\mu}a_{\mu}^+ + iAa_D^+)|0\rangle
\quad and \quad \ra{3}_0=iC|0 \rangle
\end{equation}

The lagrangian
\begin{eqnarray}
L=A_{\nu}{{\partial}_{\mu}}^2 A_{\nu} -
 A{{\partial}_{\mu}}^2A +(\alpha'+\alpha_0)
{A_{\alpha}}^2 - (\alpha'+\alpha_0) A^2 -
\nonumber \\
 C^2 - 2\sqrt{(\alpha'+\alpha_0)}CA
+C{\partial}_{\mu}A_{\mu} -A_{\mu}{\partial}_{\mu}C
\end{eqnarray}
is invariant under the following gauge transformations
\begin{eqnarray}
\delta A_{\mu}&=&{\partial}_{\mu}\lambda \\
\delta A&=&-\sqrt{\alpha'+\alpha_0}\lambda \\
\delta C&=&({\partial_{\mu}}^2 + \alpha' + \alpha_0)\lambda
\end{eqnarray}
leading on mass shell to the equations
\begin{equation}
A=C=0,\; \partial_\nu A_\nu=(\partial_\nu^2+\alpha'+\alpha_0)A_\mu=0.
\end{equation}

The corresponding formulas for $n=2$  are:
\begin{eqnarray}
\ra{1}_2&=&(A_{\mu \nu}a_{\mu}^+a_{\nu}^+ + 2 iA_{\mu}a_{\mu}^+
a_{D}^+ + Aa_{D}^+a_{D}^+)|0\rangle \\
\ra{2}_0&=&B|0 \rangle \\
\ra{3}_1&=&(iC_{\mu}a_{\mu}^+ + Ca_D^+)|0\rangle
\end{eqnarray}
\begin{eqnarray}
L&=&-2A_{\nu \rho}{\partial_{\mu}}^2A_{\nu \rho} -2(2\alpha'+\alpha_0)
{A_{\nu \rho}}^2 + 4 A_{\nu}
{\partial_{\mu}}^2A_{\nu} +
4(2\alpha'+\alpha_0)
{A_{\nu}}^2 - \nonumber\\\nonumber
&& 2 A{\partial_{\mu}}^2A -2(2\alpha'+\alpha_0)A^2  +
 B{\partial_{\mu}}^2B + (2\alpha'+\alpha_0) B^2 - C^2+{C_{\nu}}^2-
\nonumber \\&&2C_{\nu}\partial_{\mu}A_{\nu \mu}
+2A_{\nu \mu}\partial_{\mu} C_{\nu}
-2C\partial_{\mu}A_{\mu} +2A_{\mu}\partial_{\mu}C
- 4\sqrt{2\alpha'+\alpha_0}CA +
\nonumber \\
&&
4\sqrt{2\alpha'+\alpha_0}C_{\mu}A_{\mu}
+ B\partial_{\mu}C_{\mu} -
C_{\mu}\partial_{\mu}B +2 \sqrt{2\alpha'+\alpha_0}BC
\end{eqnarray}
The gauge transformations with the parameter
$|\lambda\rangle=(i\lambda_{\mu}a_{\mu}^+ + \lambda a_D^+)|0\rangle$
give:

for $\ra{1}_2$ :
\begin{eqnarray}
\delta A_{\mu \nu}&=&\frac{1}{2}(\partial_{\mu}{\lambda}_{\nu}
+\partial_{\nu}{\lambda}_{\mu}), \\
2\delta A_{\mu}&=&-\partial_{\mu}\lambda -\sqrt{2\alpha'+\alpha_0}
{\lambda}_{\mu}, \\
\delta A&=&-\sqrt{2\alpha'+\alpha_0}\lambda,
\end{eqnarray}

for $\ra{2}_0$:
\begin{equation}
\delta B = -\partial_{\mu}{\lambda}_{\mu}
-\sqrt{2\alpha'+\alpha_0}\lambda,
\end{equation}
and for $\ra{3}_1$:
\begin{eqnarray}
\delta C_{\mu}&=&({\partial_{\nu}}^2
+ 2\alpha'+\alpha_0){\lambda}_{\mu}, \\
\delta C&=&({\partial_{\nu}}^2 + 2\alpha'+\alpha_0){\lambda}.
\end{eqnarray}
After the gauge fixing the only nonvanishing field is $A_{\mu\nu}$
under conditions
\begin{equation}
(\partial_{\rho}^2+2\alpha'+\alpha_0)A_{\mu\nu}=0,\;
\partial_\nu A_{\mu\nu}=0.
\end{equation}
Such equations describe simultaneously spin 2 and spin 0 particles.

\setcounter{equation}0\section{Conclusions}
In this paper we have applied the BRST approach to the description
of the free Regge trajectory. The corresponding BRST - charge was
constructed with the help of dimensional reduction method. The 
resulting spectrum contains reducible representations of the 
Poincare group at each mass level due to luck of tracelessness 
constraint. It means that daughter trajectories also belong
to the spectrum.
The modification of the BRST approach to the single Regge
trajectory will be given elsewhere.
\vspace{1cm}

\noindent {\bf Acknowledgments.}
One of us (A.P.) would like to thank M. Vasiliev and \\C. Preitschopf
for useful discussions.
This investigation has been supported in part by the
Russian Foundation of Fundamental Research,
grants 96-02-17634 and 96-02-18126,grant MXL200,
 joint grant RFFR-DFG 96-02-00186G,
and INTAS, grant 94-2317 and grant of the
Dutch NWO organization.

\end{document}